\documentclass[
reprint, 
 superscriptaddress,
 longbibliography,
 aps,
 prb,
]{revtex4-2}
\usepackage{amsmath}
\usepackage{graphicx}% Include figure files
\usepackage{dcolumn}% Align table columns on decimal point
\usepackage{bm}% bold math
\usepackage[version=3]{mhchem} % Formula subscripts using \ce{}
\usepackage[T1]{fontenc}       % Use modern font encodings
\usepackage{amssymb}
\usepackage{txfonts}
\usepackage[usenames,dvipsnames]{xcolor}
\usepackage{xspace}
\usepackage{caption}
\usepackage{xr}
\usepackage{soul}
\usepackage{ragged2e}
\usepackage{tikz}
\usepackage [mathscr] {eucal}

\usepackage[breaklinks=true,colorlinks,citecolor=blue,linkcolor=blue,urlcolor=blue]{hyperref}

\usepackage{tikz,xcolor,hyperref}

\definecolor{lime}{HTML}{A6CE39}
\DeclareRobustCommand{\orcidicon}{
	\begin{tikzpicture}
	\draw[lime, fill=lime] (0,0) 
	circle [radius=0.16] 
	node[white] {{\fontfamily{qag}\selectfont \tiny ID}};
	\draw[white, fill=white] (-0.0625,0.095) 
	circle [radius=0.007];
	\end{tikzpicture}
	\hspace{-2mm}
}
\foreach \x in {A, ..., Z}{\expandafter\xdef\csname orcid\x\endcsname{\noexpand\href{https://orcid.org/\csname orcidauthor\x\endcsname}
			{\noexpand\orcidicon}}
}
\begin{document}

%\preprint{APS/123-QED}

\title{Strain-Induced Activation of Symmetry-Forbidden Exciton-Phonon Couplings for Enhanced Phonon-Assisted Photoluminescence in MoS$_2$ Monolayers}% Force line breaks with \\

\author{Rishabh Saraswat}
%\email{rse2022507@iiita.ac.in}
\author{Rekha Verma}
%\email{r.verma@iiita.ac.in}
\affiliation{
Department of Electronics and Communication Engineering, Indian Institute of Information Technology-Allahabad, Uttar Pradesh 211015, India}
\author{Sitangshu Bhattacharya}
%\email{sitangshu@iiita.ac.in}
\affiliation{Electronic Structure Theory Group, Department of Electronics and Communication Engineering, Indian Institute of Information Technology-Allahabad, Uttar Pradesh 211015, India}
\begin{abstract}
\noindent Phonon-assisted photoluminescence (PL) in molybdenum-based two-dimensional dichalcogenides is typically weak due to the dormant phonon coupling with optically inactive momentum-dark (intervalley) excitons, unlike in tungsten-based dichalcogenides where such processes are more prominent. Despite this inefficiency, we revisit excitons in MoS$_2$ using rigorous finite-momentum Bethe-Salpeter equation calculations to identify ways to enhance phonon-assisted recombination channels. Our ab-initio results, complemented by group-theoretic analyses, reveal that while unstrained MoS$_2$ exhibits no phonon-assisted PL emissions at cryogenic temperatures due to forbidden A$^{\prime\prime}$ phonon modes, biaxial strain opens a pathway to significantly intensify this emission by activating hole-phonon A$^{\prime}$-mediated scattering channels. By calculating allowed exciton-phonon matrix elements and scattering rates, we demonstrate how strain redistributes oscillator strengths toward radiative recombination. These findings provide a promising route to improving PL emission efficiency in various metal dichalcogenide monolayers through strain engineering and offer valuable insights for further exploration of exciton-phonon dynamics, including time-resolved spectroscopic studies.
\end{abstract}
%Exciton-driven optical absorption processes in two-dimensional (2D) materials outperform those in traditional bulk materials due to intense coulombic interactions and weak dielectric screening.                     
\maketitle
\begin{figure*}[!ht]
  \centering
  \includegraphics[width=1.00\linewidth,height=0.60\linewidth]{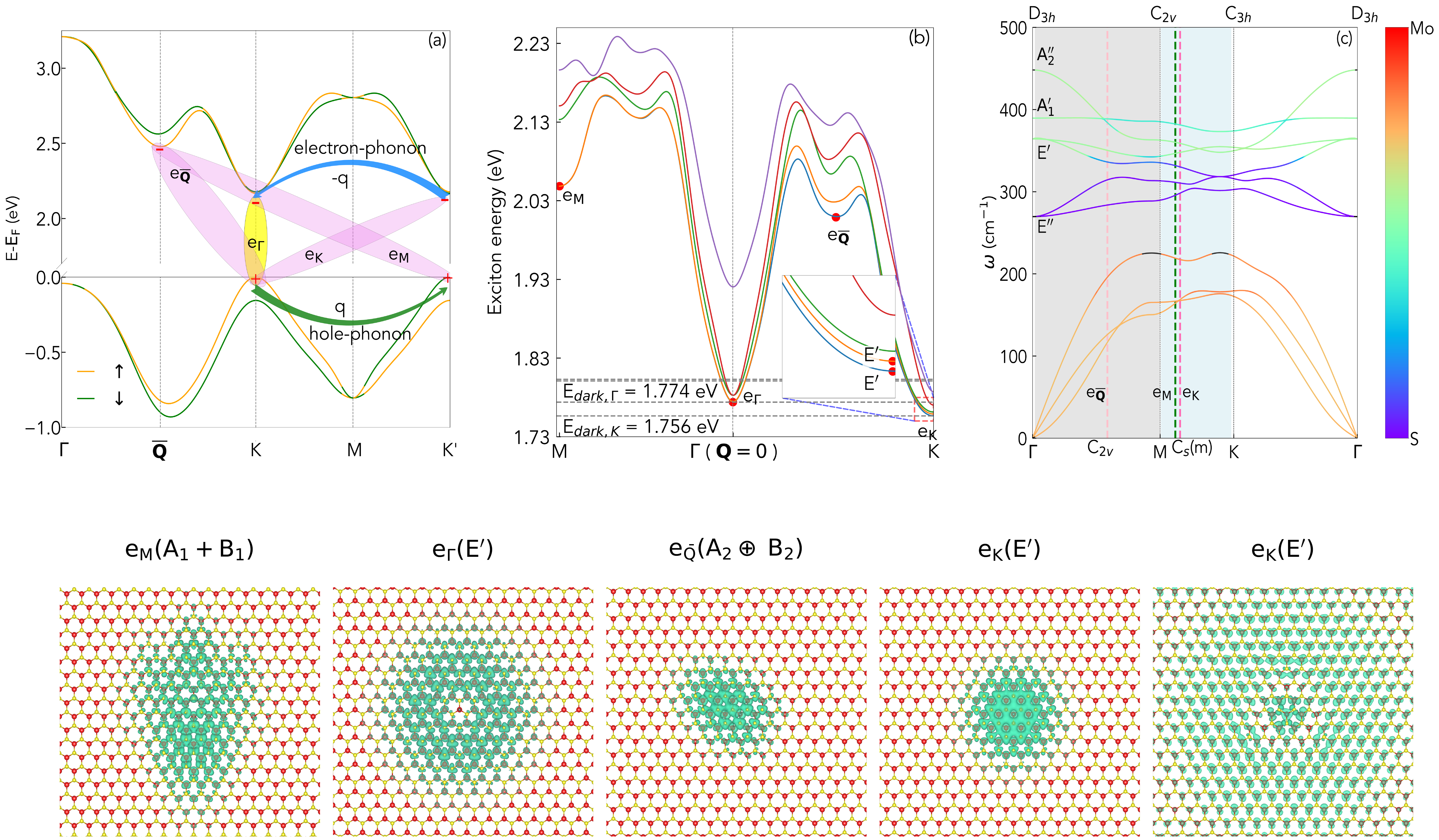}
  \caption{\justifying
  \textbf{Top panel}: (a) Spin-resolved G$_0$W$_0$ corrected electron energy dispersion in 1\% biaxially tensile strained 2D h-MoS$_2$. The intravalley excitons (labeled as \( \mathrm{e}_{\mathbf{\Gamma}} \)) are at \(\mathbf{\textbf{K}}\) and \(\mathbf{\textbf{K}}^{\prime}\).  The intervalley excitons (\( \mathrm{e}_{\mathbf{K}} \), \( \mathrm{e}_{\mathbf{M}} \), and \( \mathrm{e}_{\bar{\mathbf{Q}}} \)) are, however, momentum-forbidden. Only the top valence band is considered for holes, signifying the ground state exciton. The electron-phonon and hole-phonon scattering processes responsible for PL emissions are shown by fancy arrows. 
  (b) Corresponding excitonic energy dispersion showing the lowest five excitons along the BZ. The \(\mathbf{\Gamma}\) point (\(\textbf{Q}=0\)) corresponds to \(\mathbf{K}\) or \(\mathbf{K}^{\prime}\) in the electronic dispersion. The lowest five excitons at \(\mathbf{\Gamma}\) in ascending orders are 1 (labeled as \( \mathrm{e}_{\mathbf{\Gamma}} \)), 2, 3, 4 and 5. at Q=0. The inset magnifies the splitting of exciton energies at \(\mathbf{K}\). 
  (c) Phonon dispersion exhibiting phonon energy modes along the BZ. The individual atomic contribution is superimposed and shown by the color map. The dashed lines (\( \mathrm{e}_{\mathbf{K}} \) and \( \mathrm{e}_{\mathbf{M}} \)) in the shaded area represent the two different phonon-assisted channels for exciton recombination in (a). 
  \textbf{Bottom panel} (left to right): Excitonic probability distribution function and symmetry of the lowest exciton (shown by red circles in (b)) at momenta \(\mathbf{M}\) (\( \mathrm{e}_{\mathbf{M}} \)), \(\mathbf{\Gamma}\) (\( \mathrm{e}_{\mathbf{\Gamma}} \)), \(\mathbf{Q}_{\mathbf{\Gamma} - \mathbf{K}} \) (\( \mathrm{e}_{\bar{\mathbf{Q}}} \)) and at \(\mathbf{K}\) (\( \mathrm{e}_{\mathbf{K}} \)).}
  \label{fgr:fgr1}
\end{figure*}

%\tableofcontents
%\section{\label{sec:level1}Introduction}
%\vspace{-1.8em}
\noindent Light emission in pristine semiconductors arises from the radiative recombination of bound electron-hole pairs, known as excitons. The nature of this emission—whether direct or indirect—is dictated by the excitonic dispersion, analogous to electronic band dispersion, particularly in relation to the valley center, $\Gamma$, in the excitonic Brillouin zone (BZ) \cite{knox1963}. Intravalley excitons form at the zero of the excitonic center-of-mass momentum (\textbf{Q} = 0), which corresponds to the optical dipole limit. In this case, a direct optical gap occurs if the lowest-energy exciton is located at $\Gamma$, allowing efficient radiative recombination without additional momentum transfer, leading to intense direct photoluminescence (PL) \cite{Landsberg_1992}. In contrast, an indirect optical gap arises when the lowest exciton state lies at a finite momentum (\textbf{Q} $\neq$ 0), forming intervalley excitons, whose recombination necessitates phonon-mediated momentum transfer. While indirect processes are generally less efficient due to multiparticle interactions and selection rules, which suppress PL emission \cite{Dresselhaus2008}, bulk honeycomb (h)-boron nitride (BN) defies this limitation. Despite its indirect optical gap, exceptionally strong exciton-phonon coupling enables intense ultraviolet PL \cite{Cassabois2016}, with an efficiency surpassing even that of diamond \cite{Leonard2019}. \\
\noindent Beyond bulk materials, high-quality two-dimensional (2D) transition metal dichalcogenides (TMDs) provide an ideal platform to study spin-momentum-coupled exciton-light interactions.  These noncentrosymmetric crystals, with time-reversal symmetry, feature two inequivalent valleys (\textbf{K} and \textbf{K}$^{\prime}$) in the electronic BZ, where spin-valley locking governs exciton formation (see Fig. \ref{fgr:fgr1}(a)) \cite{WangYao2008, DiXiao2012}. Circularly polarized light with left or right handedness can selectively excite excitons in these valleys, leading to valley polarization \cite{Cao2012, Hualing2012, Mak2012, Jones2013, Srivastava2015, Echeverry2016, Deilmann2017, Malic2018, Gang2018, Mueller2018}. The relative spin alignment of conduction and valence band edges at \textbf{K} and \textbf{K}$^{\prime}$ in W- (WS$_2$, WSe$_2$) and Mo-based (MoS$_2$, MoSe$_2$) TMDs further imposes the selection criterion on exciton formation. Intravalley (\textbf{K}-\textbf{K}) excitons are bright if spin-allowed (parallel spin) and dark if spin-forbidden (antiparallel spin). In contrast, both spin-allowed and spin-forbidden intervalley (\textbf{K}-\textbf{K$^{\prime}$}) excitons are momentum-forbidden and thus optically dark \cite{Malic2018, Andrea2010, Molas2017, Robert2016, Humberto2013, Zhang_2014, Gang2018, Palummo2015, He2014}. Only bright intravalley excitons undergo radiative recombination, resulting in strong direct PL \cite{Andrea2010, Humberto2013, Zhang_2014, Gang2018, Palummo2015, He2014}. A key distinction exists between the ground-state excitons in W- and Mo-based TMDs. In W-based compounds, both intra- and intervalley ground-state excitons are degenerate and dark due to spin- and momentum-forbidden transitions \cite{Xiao_2015, Gang2018}, with dark-bright splittings of 40–50 meV \cite{Zhou2017}. In contrast, ab initio calculations predict that Mo-based TMDs host bright degenerate intra- and intervalley excitons \cite{Malic2018}, whereas experiments suggest a dark ground-state exciton with dark-bright splittings ranging from 14 to 100 meV \cite{Molas2017, Robert2020}.
\\ \noindent Excitons those are dark either by spin or momentum forbidden, do not contribute to the PL directly, unless they are activated intentionally. This has been shown for the energetically lowest lying dark intravalley spin-forbidden excitons. These excitons were observed to undergo dark to bright transition in MoS$_2$ and MoSe$_2$ \cite{Xiao-2017, Robert2020} by spin-flipping process in an external in-plane magnetic field, or due to exchange interaction \cite{Yu2014}, or by phonon assistance such as in WS$_2$ \cite{Wang-2018}, leading to valley depolarization. The recombination of momentum forbidden dark excitons at cryogenic temperatures in MoSe$_2$ has been observed with varied outcomes. In one instance, no phonon-assisted PL is reported, though there is an asymmetry in the brightest direct PL signal toward higher energies \cite{Samuel2020}. In other observations, strong exciton-phonon ($exc$-$ph$) coupling appears, assisted by longitudinal acoustic (LA) phonons at the \textbf{M} point (in the phonon BZ) \cite{Colin2017} and A$_{1}^{\prime}$ optical phonons at room temperature \cite{Donghai2021}. In contrast, WSe$_2$ consistently shows valley depolarization effects due to intense $exc$-$ph$ scattering \cite{Shoujun2017, Chellappan2018, Miyauchi2018, Samuel2020, Young2020, Funk2021, Hsiao2022}.\\
%, WangDai, ChangChen
\noindent Although phonon-assisted PL emission has garnered significant attention in 2D h-MoSe$_2$, WS$_2$, and WSe$_2$, however, such a detailed mechanism in 2D h-MoS$_2$ remains largely unexplored. Therefore, in this work, we revisit 2D h-MoS$_2$ for a comprehensive understanding of phonon-assisted indirect excitonic recombination. While MoS$_2$ direct emission around room temperatures dominates at $\mathbf{\Gamma}$ in the excitonic BZ due to thermalization, we rather show using group-theoretic analyses that the cryogenic emission due to the low-energy dark excitons are forbidden by symmetry conditions. However, by harnessing strain engineering, these dark excitons can be activated to couple with intervalley phonon mode (A$^{\prime}$)  within the \textbf{M}-\textbf{K} BZ route (see Fig. \ref{fgr:fgr1}(c)), unlocking a pathway to enhanced indirect PL emission. We demonstrate these emissions using biaxial strain in the range of -0.5$\%$ to +3$\%$, achievable under standard laboratory conditions \cite{Hong2015}. We illustrate that indirect PL emission at higher tensile strains in 2D MoS$_2$ is dominated by hole-phonon processes (see Fig. \ref{fgr:fgr1}(a)), where the hole is scattered by a phonon from \textbf{K} to \textbf{K}$^{\prime}$, resulting in a recombination at \textbf{K}$^{\prime}$. The PL intensity peaks sharply at +1$\%$ strain, with a pronounced decline above and below this range due to a reduction in $exc$-$ph$ coupling across different modes. These systematic findings can be useful as a textbook example to reveal the symmetry-driven indirect PL in 2D TMDs. The following discussions can also explain the non-monotonic PL response to biaxial strain recently observed in 2D WSe$_2$ \cite{Pablo2022}.\\ 
\noindent We employ fully relativistic, norm-conserving pseudopotentials \cite{hamann2013optimized} with core corrections for Mo (core: [Ca]; valence: 3d, 4p, 5s, and 4d) and S (core: [Ne]; valence: 3s and 3p). Ground-state and excited-state computations were performed using ab-initio codes Quantum Espresso \cite{Giannozzi2017} and Yambo \cite{Sangalli2019}, respectively. 
\begin{figure*}[!ht]
   \centering
 \includegraphics[width=1\textwidth]{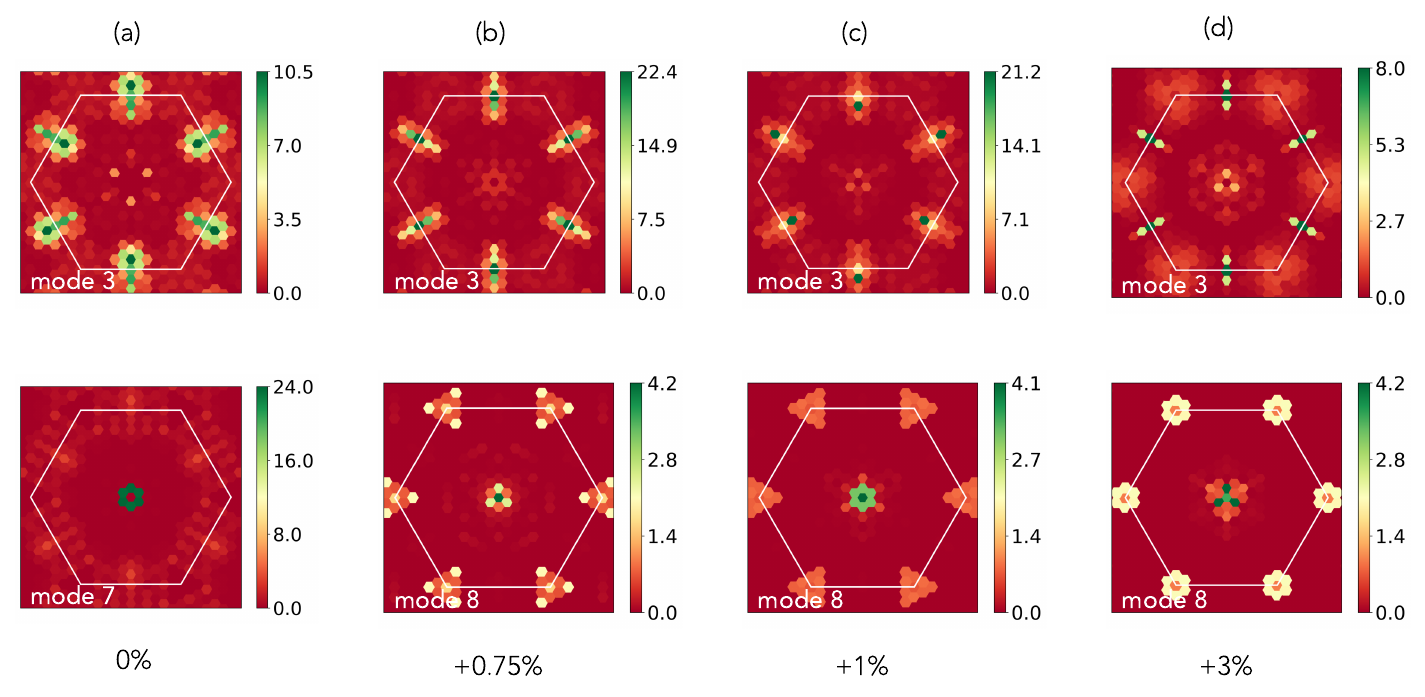}
   \caption{\justifying (a)-(c): Phonon mode A$^\prime$ resolved gauge-invariant $exc$-$ph$ coupling $\left|\mathcal{G}_{\beta\lambda,\nu}\left(\textbf{Q},\textbf{q}\right)\right|^2$ across the exciton BZ at various strains. All values are normalized to 10$^{-4}$ eV$^2$. Only the most prominent mode interaction contributing to the indirect and direct PL emission (Fig. \ref{fgr:fgr6} below) are shown. The $x$ and $y$ axes are in transferred exciton momenta. The intensity of the color-map shows the coupling strength.}  
   \label{fgr:fgr2}
 \end{figure*}
A total of 200 electronic states (26 occupied, 174 unoccupied) were sufficient to converge quasi-particle (QP) gaps, with a screening matrix cut-off of 20 Ry for both QP energies and the Bethe-Salpeter (BS) equation. Convergence tests (see Supplemental Information (SI) \cite{Supplemental}) indicate that an 18$\times$18$\times$1 grid with Coulomb cut-off techniques \cite{Guandalini2023} accurately reproduces excitonic energies within experimental ranges. Electron-phonon matrix elements (including self-energies \cite{Fan1950} with Debye-Waller corrections \cite{cannuccia2011effect}), and $exc$-$ph$ matrix elements were computed on a converged 120$\times$120$\times$1 transferred momenta fine grid. Implementation of ab-initio biaxial strain and subsequent discussion on the electronic, phonon and exciton energies including comparison with reported experiments are summarized in section I of the SI \cite{Supplemental}. In the following discussion, we focus on results at +1$\%$ biaxial strain and compare them to other strain conditions as necessary. 
% \section{\label{sec:level2}Results and Discussions}
% \vspace{-1.2em}
\\ \noindent The indirect emission process is a two-step mechanism governed by Fermi's golden rule. An exciton transitions from its initial state, $\left|\varphi_{i}\right\rangle$, to a final state, $\left|\varphi_{f}\right\rangle$, via an intermediate state, $\left|\varphi_{t}\right\rangle$. This involves a light coupling $\left(\hat{\varsigma}\right)$ and phonon coupling $\left(\hat{g_{\textbf{q}}}\right)$, which may occur in either order. The entire process can then be described by $\left\langle \varphi_{f}\left|\hat{g}^{\dagger}_\textbf{q}\hat{\varphi}\right|\varphi_{i}\right\rangle = \left\langle \varphi_{f}\left|\hat{g}^{\dagger}_\textbf{q}\right|\varphi_{t}\right\rangle \left\langle \varphi_{t}\left|\hat{\varsigma}\right|\varphi_{i}\right\rangle$. The light-field polarization direction also impose condition on the coupling of phonon modes with excitons via the symmetry of states that are excited, aligning with the selection rules dictated by the crystal's point group symmetry. Such cases are reported for bulk h-BN \cite{PaleariVarsano}, where in-plane and out-of-plane light-field polarization selectively allows LA/TA and Z phonon modes coupling with excitons \cite{Fulvio2019, Vuong2017} and is found effective in overcoming limitations of indirect emission in WS$_2$ \cite{Yiming2024, Tamaghna2024} and WSe$_2$ \cite{Ozgur2018} by strain engineering.\\
For a 2D MoS$_2$, the D$_{3h}$ point group symmetry includes a fully symmetric (even) irreducible representation, $A_{1}^{\prime}$. The in-plane (two-dimensional) components, $x$ and $y$, transform as the $E^{\prime}$ representation, enabling mixing and transitions allowed by symmetry. The out-of-plane $z$ component transforms as $A_{1}^{\prime}$, allowing additional transitions. Consequently, the overall transformation of the dipole operator can be expressed as $E^{\prime}\left[x,y\right] + A_{1}^{\prime}\left[z\right]$. This implies that when an incoming plane-polarized light field (used commonly in pump-probe experiment), represented by $E^{\prime}\left[x,y\right]$, interacts with a ground state of symmetry $A_{1}^{\prime}$, the resulting tensor product becomes $A_{1}^{\prime} \otimes E^{\prime} = E^{\prime}$. This corresponds to the symmetry of the energetically lowest doubly degenerate dark excitons ($\mathrm{e}_{\mathbf{\Gamma}}$) at the $\mathbf{\Gamma}$ point. \\
% \textcolor{blue}{We would like to clarify the exciton labeling scheme here. The exciton labels (e.g., 1, 2, 3, 4, 5) are defined based on ascending exciton energies at specific points in momentum space, without factoring in degeneracies. At $\textbf{Q} = 0$, the lowest-energy states appear in pairs of degenerate states, with states 1 and 2, forming degenerate pairs. State 3 corresponds to the bright A exciton, degenerate with state 4 (which remains dark). State 5 is a higher-energy dark exciton without a degenerate partner within the chosen energy range.}
To analyze the symmetry of the intervalley excitons, we performed a rigorous finite-momentum BS equation calculation for the lowest five excitonic energies and states across the entire BZ. Momentum conservation indicates that the intervalley exciton $\mathrm{e}_{\mathbf{K}}$ in Fig. \ref{fgr:fgr1}(a) corresponds to the exciton at the \textbf{K} point in the exciton dispersion shown in Fig. \ref{fgr:fgr1}(b). Similarly, exciton $\mathrm{e}_{\mathbf{M}}$ and $\mathrm{e}_{\mathbf{\bar{Q}}}$ are located at the \textbf{M} and about $\frac{1}{2}\left|\mathbf{\Gamma}-\textbf{K}\right|$ point respectively. The degeneracy at $\mathbf{\Gamma}$ is lifted due to spin-orbit coupling along the $\mathbf{\Gamma}$-\textbf{K} direction. At the \textbf{K} point, the point group symmetry reduces to C$_\mathrm{3h}$. The two-dimensional irreducible representation, $E^{\prime}$, transforms as the one-dimensional $E^{\prime}$ in C$_\mathrm{3h}$, which is even with respect to the mirror symmetry $\sigma_h$ and highlights that the charge densities are predominantly localized at the hole sites. This corresponds to the symmetry of the lowest two excitons $\mathrm{e}_{\mathbf{K}}$ at \textbf{K}.  Similarly, the point group symmetry at the \textbf{M} towards $\mathbf{\Gamma}$ reduces to C$_\mathrm{2v}$, resulting in the double-point group irreducible representation $\Gamma_{5}$, describing the degenerate excitons $\mathrm{e}_{\mathbf{M}}$ at \textbf{M}. Additionally, the exciton $\mathrm{e}_{\mathbf{\bar{Q}}}$ is odd with respect to both $\sigma_{v}$ and $\sigma_{v'}$ rotations within the C$_{2v}$ group and as the degeneracy is lifted at this point, the split excitons have either A$_2$ or B$_2$ symmetry. We emphasize that biaxial strain in the 2D system does not alter the crystalline point group symmetry; thus, the symmetry representations remain invariant under applied strain. These symmetries are illustrated in the bottom panel of Fig. \ref{fgr:fgr1}. \\
\noindent Similar to the electron-phonon matrix elements \cite{Giustino2017, Lechifflart2023}, 
\begin{equation} 
g_{m n,\nu}(\textbf{k},\textbf{q}) = \langle m\textbf{k} |\Delta V_{\nu \textbf{q}}| n\textbf{k}-\textbf{q} \rangle 
\label{eqn:eqn1}
\end{equation}
\noindent where $g_{mn,\nu}(\textbf{k},\textbf{q})$ denotes the electronic scattering probability amplitude for a transition from an initial Bloch state $\left|n\textbf{k}-\textbf{q}\right\rangle$ to a final state $\left|m\textbf{k}\right\rangle$  via phonon absorption or emission and $\Delta V_{\nu \textbf{q}}$ represents the Kohn-Sham potential perturbed by phonons, the $exc$-$ph$ matrix elements in the excitonic basis can be written as \cite{Lechifflart2023, Marini2024, Paleari2019}
\begin{multline}
\mathcal{G}_{\beta\lambda,\nu}\left(\mathbf{Q},\mathbf{q}\right)=\sum_{\upsilon,\upsilon^{\prime},c,c^{\prime},\mathbf{k}}A_{\lambda,\mathbf{Q}}^{\upsilon,c,\boldsymbol{\mathbf{k}}}\left[g_{\upsilon\upsilon^{\prime},\nu}\left(\mathbf{k}-\mathbf{Q},\mathbf{q}\right)\delta_{c,c^{\prime}}\right]A_{\beta,\mathbf{Q+q}}^{\upsilon^{\prime},c^{\prime},\boldsymbol{\mathbf{k^{\ast}}}} \\
-\sum_{\upsilon,\upsilon^{\prime},c,c^{\prime},\mathbf{k}}A_{\lambda,\mathbf{Q}}^{\upsilon,c,\boldsymbol{\mathbf{k}}}\left[g_{c^{\prime}c,\nu}^{\ast}\left(\mathbf{k}+\mathbf{q},\mathbf{q}\right)\delta_{\upsilon,\upsilon^{\prime}}\right]A_{\beta,\mathbf{Q+q}}^{\upsilon^{\prime},c^{\prime},\boldsymbol{\mathbf{k+q^{\ast}}}}
\label{eqn:eqn2}
\end{multline}
where \(\mathcal{G}_{\beta\lambda,\nu}(\mathbf{Q},\mathbf{q}\)) now denotes the probability amplitude for scattering of an exciton $\lambda$ into exciton $\beta$, mediated by absorption or emission of a phonon characterized by state $\left|\nu \textbf{q}\right\rangle $. The coupling strength is defined as $\left|\mathcal{G}_{\beta\lambda,\nu}\left(\textbf{Q},\textbf{q}\right)\right|^{2}$.
\begin{figure}[!ht]
   \centering
 \includegraphics[width=1\columnwidth]{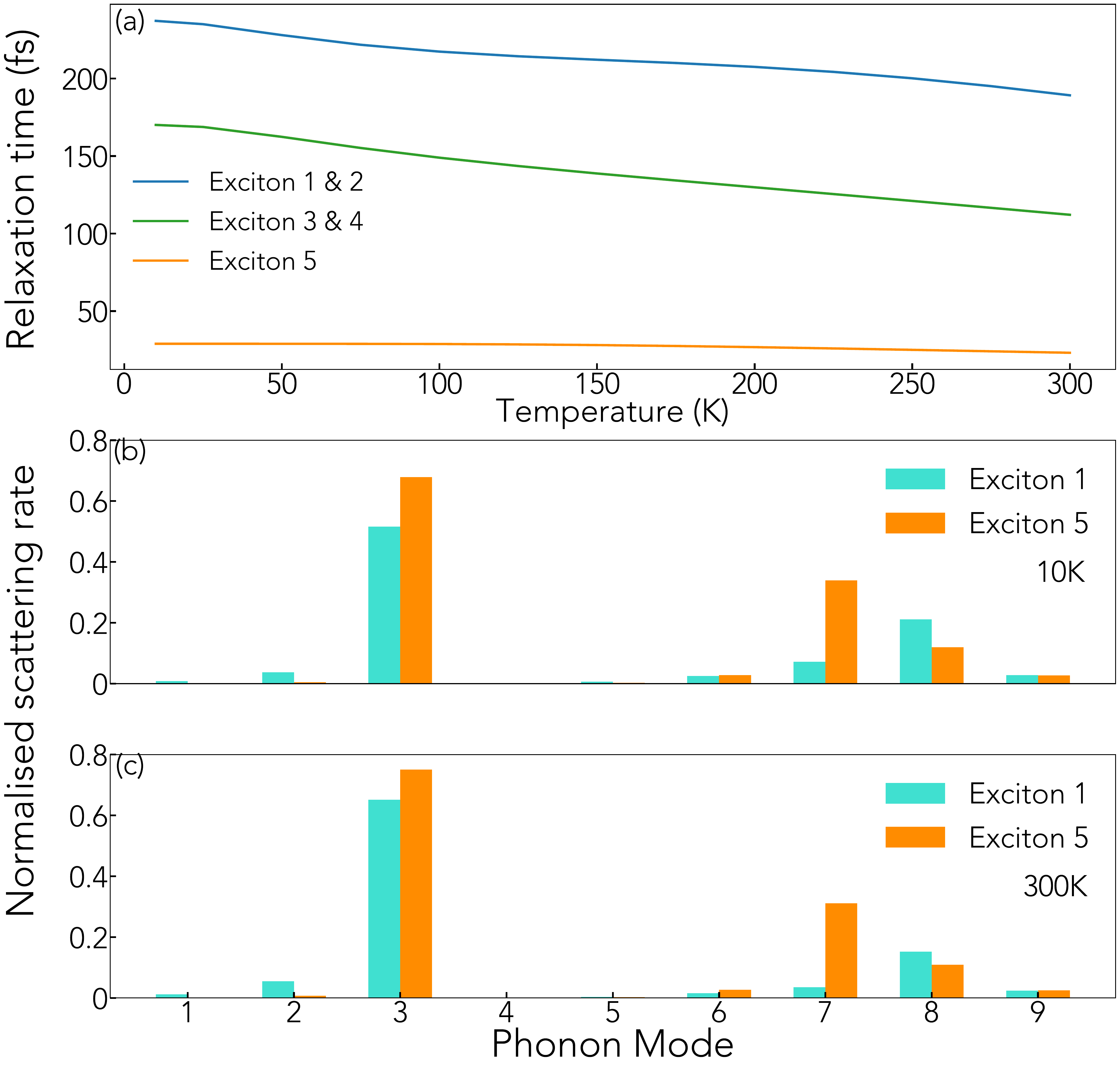}
   \caption{\justifying (a) Relaxation time at +1$\%$ strain for the lowest five excitons: 1-5 (see Fig. \ref{fgr:fgr1}(b)) at \textbf{Q} = 0 $\left(\Gamma\right)$ as function of excitonic temperature $\left(T_{exc}\right)$. Phonon mode resolved normalized scattering rates of excitons 1 and 5 at (b) 10 K and (c) 300 K.}  
   \label{fgr:fgr3}
 \end{figure}
The $exc$-$ph$ interaction in Eqn. (\ref{eqn:eqn2}) can be interpreted as a quantum superposition of electron and hole scattering events with phonons, with each event weighted by the exciton wavefunction represented in the transition basis \cite{Chen2020}. After incorporating the long and short range Coulomb Hartree potential (with reciprocal lattice vectors \textbf{G}=0) in the finite-BS equation calculation and obtaining the phonon energies from density functional perturbation theory (DFPT), we computed the $exc$-$ph$ coupling strength for the lowest exciton. \\
We highlight here the prominent phonon modes that strongly interact with the lowest indirect exciton. By analyzing the transferred momentum between the larger and reduced point group symmetries (Fig.~\ref{fgr:fgr1}(c)), we identify intervalley modes between \textbf{M} and \textbf{K} with C$_s$(m) symmetry as key contributors. These modes belong to branches characterized by A$^{\prime}$ (even under $\sigma_{h}$) or A$^{\prime\prime}$ (odd under $\sigma_{h}$), which drive both indirect and direct scattering processes. Using the compatibility table \cite{Dresselhaus2008}, we find that the $E^{\prime}$ exciton at \textbf{K} reduces to $A^{\prime}$ in C$_s$(m), allowing symmetric coupling. In contrast, when reducing to $A^{\prime\prime}$, the antisymmetric representation results in forbidden interactions. Consequently, modes 1, 4, 5, and 9, which are A$^{\prime\prime}$, do not contribute to indirect emission at any strain. This contrasts with excitons at $\mathbf{\Gamma}$, where all modes are symmetry-allowed to couple. Figure~\ref{fgr:fgr2} illustrates these interactions under various strain levels. The quantity shown represents the average coupling strength for transitions involving the low-energy excitonic states, specifically \((\beta,\lambda) = (1,1), (1,2), (1,3), (1,4),\) and \((1,5)\). This averaging procedure ensures that the presented coupling strength characterizes the phonon-mediated interactions primarily among these lowest-energy exciton states. For instance, the subplot (a) shows for the unstrained case, where only significant allowed modes are 3 and 7 that would exhibit noticeable interaction with the lowest exciton between $\mathbf{\Gamma}$-\textbf{M} and around $\mathbf{\Gamma}$ respectively. While other modes are less significant, the subplot demonstrates that coupling between the $\mathbf{\Gamma}$ and \textbf{M} points is primarily driven by mode 3. However, mode 7 exhibits stronger coupling, predominantly interacting with excitons localized near $\mathbf{\Gamma}$. Subplot (b) shows the $exc$-$ph$ coupling strength at +0.75$\%$ strain for the lowest exciton interacting with phonon mode 3 and 8 with mode 3 having similar strength compared to the previous case.
\begin{table}[h!]
\centering
\resizebox{\columnwidth}{!}{%
\begin{tabular}{|c|c|c|}
\hline
Reduction to & Mode: $A^{\prime}$ ($C_{s}(m)$) & Mode: $A^{\prime\prime}$ ($C_{s}(m)$) \\
\hline
Exciton: $E^{\prime}$ at \textbf{K} ($C_{3h}$) & $A^{\prime}$ &  0 \\
% Exciton: $E^{\prime}$ at $\Gamma$ ($D_{3h}$) & $A^{\prime}$ & $A^{\prime\prime}$ \\
Exciton: $\Gamma_5$ at $\mathbf{M}$ ($C_{2v}$) & $A^{\prime} \oplus A^{\prime\prime}$ & $A^{\prime} \oplus A^{\prime\prime}$ \\
\hline
\end{tabular}%
}
\caption{\justifying Allowed exciton (lowest) and phonon mode interactions in various point groups. The section II of the SI \cite{Supplemental} contains detailed discussion on our group-theoretic results.}
\label{tble:tbl1}
\end{table}
\\ \noindent At higher strain levels (\(+1\%\) and \(+3\%\)), as shown in subplots (c) and (d), the coupling strength increases significantly compared to the respective modes (see Figs. S17-S23 in the SI \cite{Supplemental} for plots of intermediate strain values). At \(+1\%\), modes 3 and 8 strongly interact with the lowest exciton near \(\mathbf{M}\) and around \(\mathbf{\Gamma}\), respectively. This observation aligns with the exciton dispersion in Fig.~\ref{fgr:fgr1}(b), which suggests phonon-assisted exciton scattering in these states. Additionally, mode 8 shows substantial interaction with the exciton near the \(\mathbf{K}\) and \(\mathbf{K}^{\prime}\) points, indicating strong scattering in these regions. At \(+3\%\), the interaction distribution for mode 3 becomes more widespread across all high-symmetry points. However for mode 8, the interaction strength significantly drop-out but becomes increasingly localized around \(\mathbf{\Gamma}\), \(\mathbf{K}\), and \(\mathbf{K}^{\prime}\) points, reflecting a shift in the coupling dynamics. 
\begin{figure}[!ht]
\centering
\includegraphics[width=1\columnwidth]{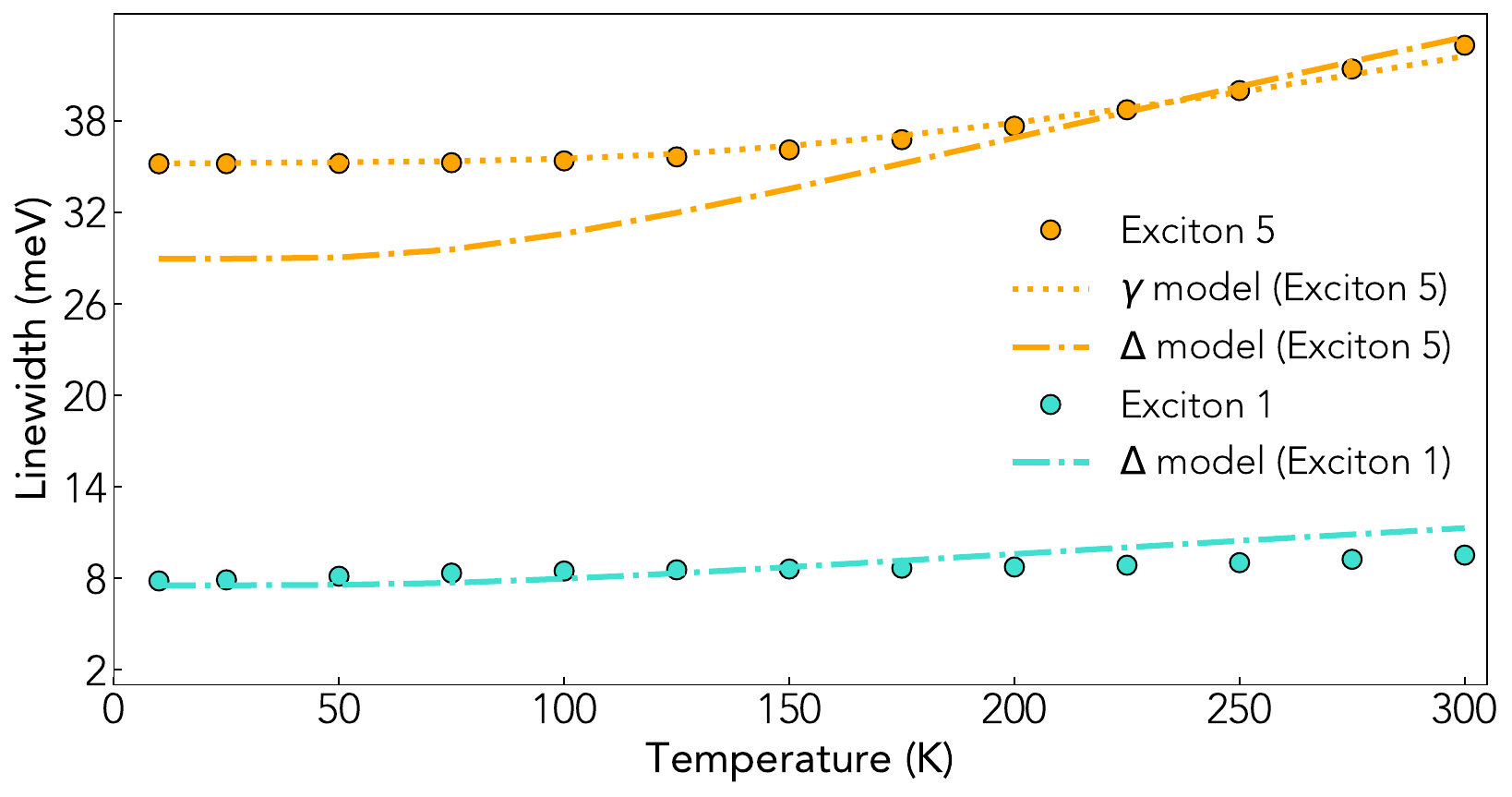}
\caption{\justifying Linewidth for exciton 1 and 5 at $\mathbf{\Gamma}$ as function of $T_{exc}$ at +1$\%$ strain. Symbols are the ab-initio result where as broken lines are fitted from the appropriate linewidth models.}  
\label{fgr:fgr4}
\end{figure}
\\ In Fig. \ref{fgr:fgr3}, we compute the scattering rate for the first five excitons at  \textbf{Q} = 0 (excitons 1 \& 2 are degenerate dark, exciton 3 is bright, degenerate with the dark exciton 4, and exciton 5 is again dark) by evaluating the imaginary part of the \textit{exc}-\textit{ph} coupling self-energy~\cite{Lechifflart2023}
\begin{multline}
\Xi_{\lambda}\left(\mathbf{Q}=0;\omega\right)=\frac{1}{\Omega_{BZ}}\sum_{\mathbf{q},\nu\beta}\mathcal{G}_{\beta\lambda,\nu}\left(\textbf{Q}=0,\mathbf{q}\right)\mathcal{G}_{\beta\lambda,\nu}^{\ast}\left(\textbf{Q}=0,\mathbf{q}\right)\\\times\left[\frac{1+n_{\mathbf{q},\nu}}{\omega-E_{\mathbf{q},\beta}+\omega_{\mathbf{q},\nu}+i\eta}+\frac{n_{\mathbf{q},\nu}}{\omega-E_{\mathbf{q},\beta}-\omega_{\mathbf{q},\nu}+i\eta}\right]
\label{self_energy}
\end{multline}
where $\Omega_{\mathrm{BZ}}$ is the BZ volume and $T_{exc}$ goes into the Boltzmann factor \cite{Fulvio2019}. In Eqn. (\ref{self_energy}), we neglect the temperature-dependent excitonic occupation factors, as they are negligibly small compared to phononic occupations. Additionally, we consider only the diagonal part of the self-energy. Under unstrained conditions, both our methodology and computed linewidth values align well with recent findings by Chan et al. \cite{Chan2023}. To verify this, we directly compare our calculated linewidths with their results in Fig. 2(b) of the SI \cite{Supplemental}. Our values closely match their linewidths, which are evaluated from the diagonal elements of the imaginary part of the \textit{exc}-\textit{ph} self-energy. Notable deviations appear only at higher temperatures, though these remain relatively small (on the order of a few meV).\\
% \\ \noindent Using the coupling strengths, we find the scattering rates for each of the five excitons at $\Gamma$ in Fig. \ref{fgr:fgr3}. The $exc$-$ph$ scattering rate can be computed from \cite{Chen2020} 
% \noindent
% \begin{multline}
% \Gamma_{n\textbf{Q}}^{\text{ex-ph}}(T) = \frac{2\pi}{\hslash} \frac{1}{\mathcal{N}_{\textbf{q}}} \sum_{nm\nu} |\mathcal{G}_{nm\nu}(\textbf{Q},\textbf{q})|^{2}\times \biggl[(N_{\nu \textbf{q}}\pm \\ 1/2 \pm F_{m\mathbf{\textbf{Q}+\textbf{q}}} + 1/2)\times \delta(E_{n\mathbf{Q}} - E^{\prime}_{m\mathbf{Q+q}} \pm \hslash\omega_{\nu \textbf{q}}) \biggr]
% \end{multline}
% where $N_{\nu \textbf{q}}$ and $F_{m\mathbf{Q+q}}$ represent the phonon and excitonic occupation factors, respectively, and $\mathcal{N}_{\textbf{q}}$ is the total number of discrete phonon points. 
The relaxation time is then given as the inverse of the scattering rate, which is plotted in Fig. ~\ref{fgr:fgr3}(a)-(c).  
From these plots, we see the general trend of increment of the scattering rate with temperature for all excitons \cite{Hsiao2022}, albeit with varying slopes. This indicates that higher temperatures lead to stronger \textit{exc}-\textit{ph} interactions across the spectrum.  
To gain further insight, we analyzed the contributions of individual phonon modes to the scattering rate at cryogenic (10 K) and room temperature (300 K) for the two excitons: exciton 1 and exciton 5 at $\mathbf{\Gamma}$. The contributions are normalized to the total scattering rate at their respective temperatures. The impact of modes 3 and 8 on the scattering of the exciton 1 is evident and is attributed to the $exc$-$ph$ interaction at both 10 K and 300 K. In contrast, for the exciton 5, modes 3 and 7 are the primary contributors at these temperatures. 

\noindent Figure~\ref{fgr:fgr4} illustrates the linewidth ($\Delta$) as a function of temperature for excitons~1 
and~5 at $\mathbf{\Gamma}$. 
For exciton 1, the $\Delta$ is interpreted to fall under the strong-coupling regime following Toyozawa's equation \cite{Toyozawa1953, Vuong2017a}:
\begin{equation} \Delta=\sqrt{\Delta_{\mathrm{A}}^{2}+\Delta_{\mathrm{O}}^{2}} \end{equation}
where $\Delta_{\mathrm{A}}^{2}=S_{\mathrm{A}}E_{\mathrm{A}}\mathrm{coth}\left(\frac{E_{\mathrm{A}}}{2k_{B}T}\right)$ represents the broadening due to acoustic phonons, and $\Delta_{\mathrm{O}}^{2}=S_{\mathrm{O}}E_{\mathrm{O}}\left[\mathrm{exp}\left(\frac{E_{\mathrm{O}}}{2k_{B}T}\right)-1\right]^{-1}$ accounts for the broadening caused by optical phonons. Here, $S_{\mathrm{A}}$ and $S_{\mathrm{O}}$ are the fitted $exc$-$ph$ coupling strengths for acoustic and optical phonons, respectively, and $E_{\mathrm{A}}$ and $E_{\mathrm{O}}$ are their mean phonon energies.
\begin{figure}[!ht]
  \centering
  \includegraphics[width=1.00\linewidth]{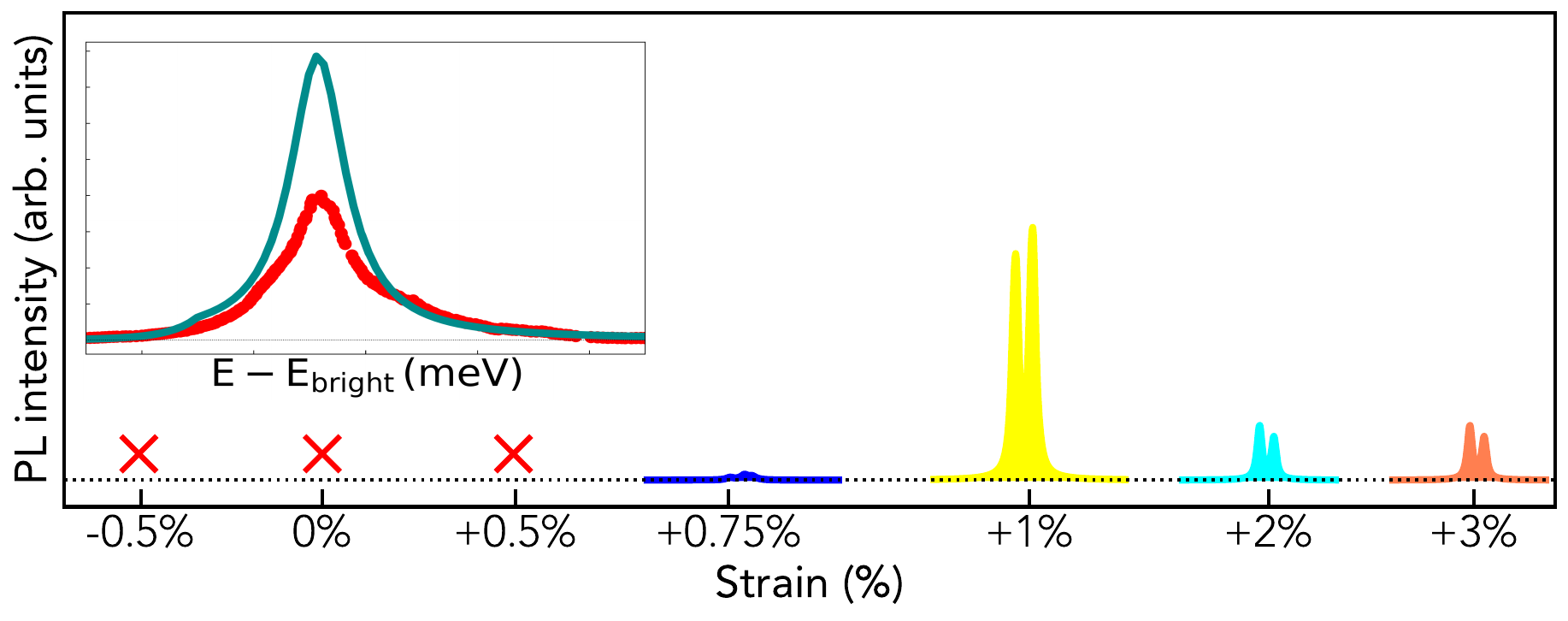}
  \caption{\justifying Lowest energy phonon replicas at 10 K for various strains. ''The $\times$`` symbol denotes a forbidden indirect PL due to symmetry. The intensity scale is absolute. The inset shows the experimental cryogenic direct PL (symbols) from \cite{Robert2020} for unstrained 2D MoS$_2$. The solid line is our ab-initio result. } 
  \label{fgr:fgr5}
 \end{figure}
%The results for the intermediate strain values such as at +0.75$\%$ and +2$\%$ are shown in the SI \cite{Supplemental}.\\
For exciton 1 under +1$\%$ strain, we fit the data using the parameters $S_{\mathrm{A}}$ = 2.27 meV, $S_{\mathrm{O}}$ = 0.1 meV, $E_{\mathrm{A}}$ = 25 meV, and $E_{\mathrm{O}}$ = 48.1 meV. In contrast, exciton 5 exhibits weak coupling at lower temperatures, which is modeled using 
\begin{equation} \gamma=\gamma_{0}+aT+b\left[\mathrm{exp}\left(\frac{E_{\mathrm{O}}}{k_{B}T}\right)-1\right]^{-1} \end{equation}
where $a$ (= 2 meV/K) and $b$ (= 35 meV) represent the contributions from acoustic and optical phonon coupling, respectively, and $\gamma_{0}$ (= 35.1 meV) is a temperature-independent offset. For exciton 5, weak coupling dominates up to approximately 225 K. Beyond this temperature, the broadening transitions to being governed by strong coupling. However, it is noted that the linewidth roll-off rates remain relatively modest.
We resolved that the modes 3 and 8 for exciton 1 while 3 and 7 (\(A^\prime\) for exciton 5,  are responsible to account such coupling in $\Delta$.\\
\noindent To compute the PL intensity, we employ the van Roosbroeck-Shockley (RS) relation \cite{Lechifflart2023}
\begin{multline}
I^{PL}(\omega) = \operatorname{Im} \sum_{\lambda} \frac{|T_{\lambda}|^2}{\pi^2 \hbar c^3} 
\Bigg[
\omega^3 n_r(\omega) \frac{1 - R_{\lambda}}{\omega - E_{\lambda} + i \eta} 
e^{-\frac{E_{\lambda} - E_{min}}{k T_{exc}}} \\
+ \sum_{\mu\, \beta\, \mathbf{q}} 
\omega (\omega \mp 2 \omega_{\mathbf{q} \mu})^2 n_r(\omega) 
\left| \mathcal{D}^{\pm}_{\beta \lambda, \mathbf{q} \mu} \right|^2 \\
\times \frac{\frac{1}{2} \pm \frac{1}{2} + n_{\mathbf{q}, \mu}}{\omega - (E_{\mathbf{q}, \beta} \mp \omega_{\mathbf{q} \mu}) + i \eta}
e^{-\frac{E_{\mathbf{q}, \beta} - E_{min}}{k T_{exc}}}
\Bigg]
\label{PL}
\end{multline}
\noindent where $|T_{\lambda}|^2$ are the excitonic dipoles, $n_r(\omega)$ is the refractive index, $R_{\lambda}$ is the renormalisation factor that measures the amount of spectral weight transferred to the satellites, $E_{\lambda}$ are the excitonic energies, $E_{min}$ is the minimum of the exciton dispersion, $\left| \mathcal{D}^{\pm}_{\beta \lambda, \mathbf{q} \nu} \right|^2$ are the phonon-assisted coupling strengths and $E_{\mathbf{q}, \beta}$ are the finite-momentum exciton energies.
In our calculations, we incorporate a finite excitonic temperature which is used in all the temperature dependent plots. %Similar to bulk h-BN, we set $T_{exc}$ to be twice the bosonic temperature $T$ \cite{Lechifflart2023}, where the bosonic temperature was found to have a negligible impact on the resulting PL spectrum. 
We understand that this is an approximation for our case and the actual result will depend on the careful experiment between the bosonic and exciton temperatures as shown in Cassabois et al. \cite{Cassabois2016, Lechifflart2023}. Unfortunately, we could not find such experimental data for MoS$_2$ to our best knowledge. Despite this, our PL and linewidth results for unstrained MoS$_2$ seems to be in strong agreement with experiments (see Fig. 2(b) in the SI and the inset in Fig. \ref{fgr:fgr5}). Additionally, a damping factor of 2 meV is applied to account for the $exc$-$ph$ self-energy, and five virtual exciton states are included in the $exc$-$ph$ scattering process.\\
We note here that due to the presence of two distinct frequency components in the exciton-phonon Hamiltonian—originating from electron-electron and electron-phonon interactions—phonon-assisted PL cannot be obtained directly by solving this Hamiltonian. Instead, the PL spectrum is derived using the Green’s function formalism, specifically through a first-order expansion of the finite-momentum exciton propagator with the dynamical exciton-phonon kernel in a Dyson-like equation. To achieve this, one should first compute the exciton propagator by solving the static finite-momentum BS equation in the absence of lattice vibrations. Next, using DFPT, the electron-phonon matrix elements are evaluated at the same \textbf{k} and \textbf{q} grids, ensuring mandatory convergence tests. These elements are then used to construct the exciton-phonon matrix elements and the corresponding self-energies within the framework of many-body perturbation theory \cite{Mahan2014}, maintaining consistency in the transferred \textbf{Q} and \textbf{q} grids with additional convergence checks. The excitonic response is subsequently obtained by expanding the Dyson-like equation up to first order. Therefore, the electronic band renormalization induced by phonons does not enter the finite-momentum excitonic BS Hamiltonian.
\begin{figure}[!ht]
  \centering
  \includegraphics[width=1\linewidth]{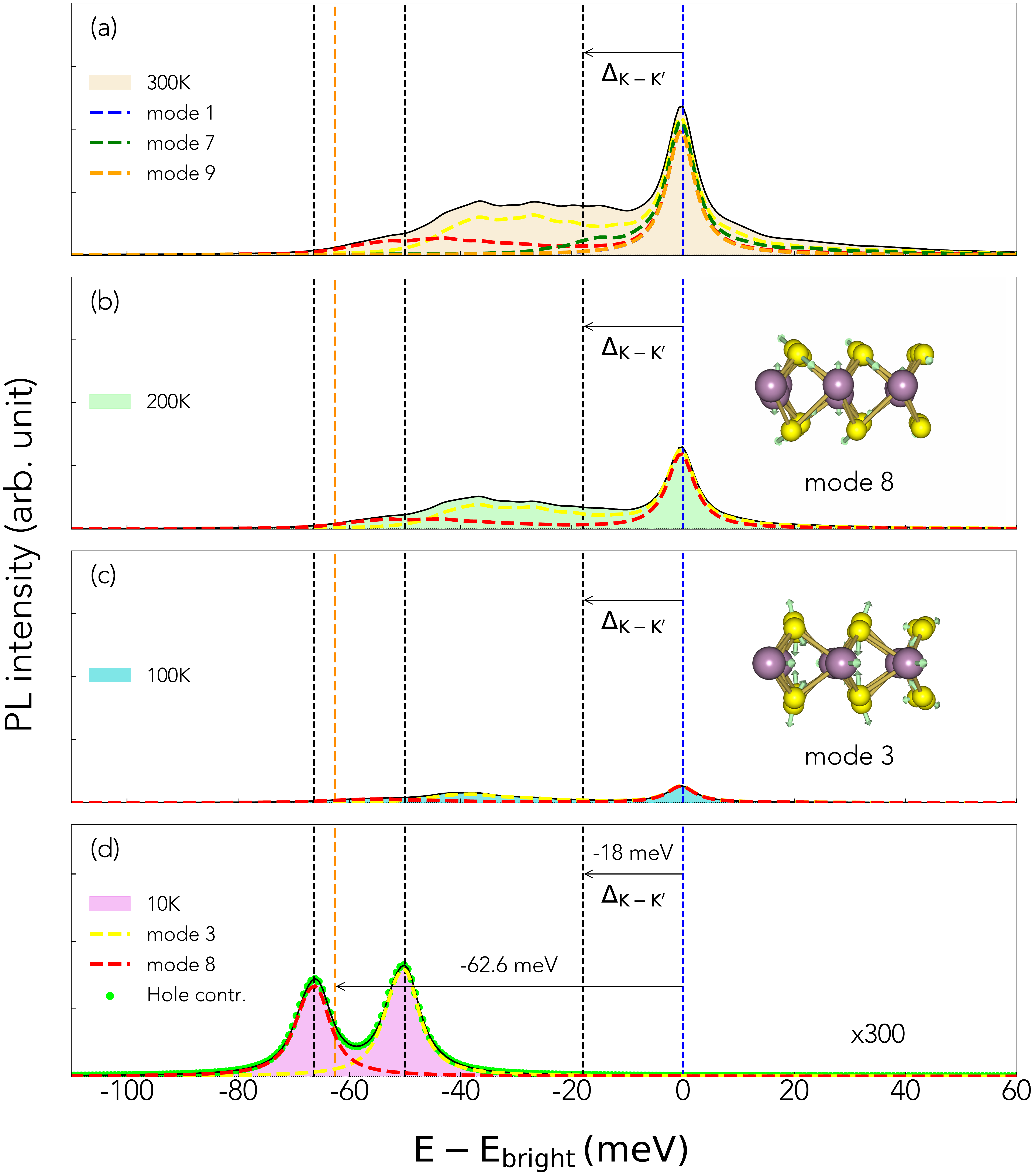}
  \caption{\justifying PL intensities with increasing $T_{exc}$ at +1$\%$ strain demonstrating phonon replicas for the modes 3 and 8. The energies are measured as shift from the first bright exciton at $\Gamma$. The inset in (b) and (c) shows the atomic vibrations for the modes 3 and 8. The dotted symbols in (d) represent the hole-phonon process in the PL intensity. } 
  \label{fgr:fgr6}
 \end{figure}
 \begin{figure*}[!ht]
  \centering
  \includegraphics[width=1.00\linewidth,height=0.20\linewidth]{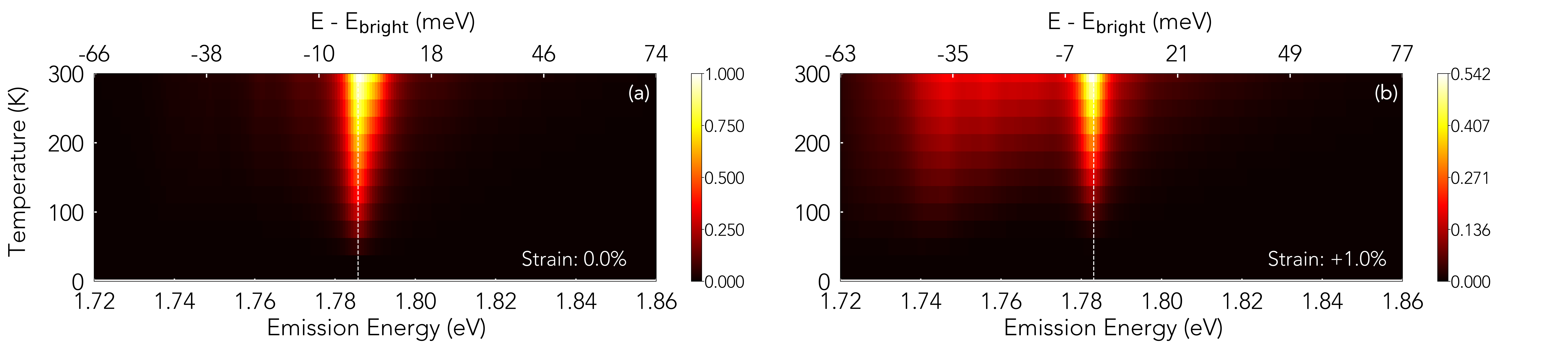}
  \caption{\justifying Normalized PL intensities with varying temperatures at unstrained and +1$\%$ strain. The vertical dashed line in all cases shows the direct bright recombination.} 
  \label{fgr:fgr7}
 \end{figure*}
To ensure our analyses are on right track, we confirmed our excitonic dispersion to be in excellent agreement with Wu et. al. \cite{Fengcheng2015} at unstrained condition. Figure \ref{fgr:fgr5} illustrates the lowest-energy PL intensity peaks under various strain conditions. Our ab-initio results for direct PL from unstrained MoS$_2$ monolayer show excellent agreement with experimental findings (inset in Fig. \ref{fgr:fgr5}) \cite{Robert2020}. We observe the emergence of a doublet structure, reaching its maximum at +1$\%$ strain driven by intense $exc$-$ph$ interactions. This is in contrast to +0.75$\%$, where the PL is only due to the dominant mode 8. To highlight this non-monotonic behavior, we present the exciton thermalization process in Fig. \ref{fgr:fgr6}(a)-(d). Similar to Malic et al. \cite{Samuel2020} we demonstrate our PL emission energies with respect to the first bright exciton energy E$_{\mathrm{bright}}$ at $\mathbf{\Gamma}$ in the excitonic BZ. Consequently, all emissions to the left of the zero-energy line correspond to indirect processes. Our results can provide an insight to the recently reported PL emission by López et. al. \cite{Pablo2022} for WSe${_2}$. While López et. al. attribute their similar non-monotonic PL variation (see Fig. 3 in their work) to the energetic resonance of indirect excitons with defect-related states, we believe that contributions from intense symmetry-allowed $exc$-$ph$ interactions, such as highlighted in this work, cannot be overlooked. \\
\noindent At cryogenic temperatures, as shown in Fig.~\ref{fgr:fgr6}(d), the emission lines exhibit two prominent phonon replicas. These Gaussian-type replicas, located approximately 50 and 70 meV below the bright line, are attributed to intervalley acoustic and optical modes 3 and 8, respectively, along the \textbf{M}-\textbf{K} direction (see the dashed line in Fig.~\ref{fgr:fgr1}(c) corresponding to the e$_2$ exciton). Mode 3 originates from in-plane vibrations of Mo atoms (shear mode), while mode 8 corresponds to out-of-plane vibrations. The PL associated with the optical mode appears at lower energy than that of the acoustic mode. To discern whether the observed peaks arise from electron-phonon or hole-phonon scattering processes (see Eqn.~(\ref{PL})), we evaluated the PL intensity separately for their contributions. Our analysis reveals that the lower-energy PL peak is primarily driven by hole-phonon scattering, where optical phonons facilitate hole scattering from \textbf{K} to \textbf{K$^{\prime}$}, leading to recombination at \textbf{K$^{\prime}$}. The strong exciton-phonon interaction responsible for this recombination is illustrated by the coupling matrix $\left|\mathcal{G}(\textbf{Q},\textbf{q})\right|^2$ in Fig.~\ref{fgr:fgr2}(c) and schematically depicted in Fig.~\ref{fgr:fgr1}(a). This behavior remains consistent at higher strains, exemplifying exciton scattering at \textbf{K} and \textbf{K$^\prime$} as a transition from a no-phonon to a hole-phonon process.\\
\noindent The next higher-energy peak, around 50 meV, corresponds to excitons near $\mathbf{\Gamma}$ along the $\mathbf{\Gamma}$-\textbf{M} direction, which involve small acoustic phonon energies for scattering. Excitons $\mathrm{e}_{\mathbf{M}}$ and $\mathrm{e}_{\mathbf{\bar{Q}}}$ scatter within themselves using intervalley optical phonon mode 8 ($\sim$42 meV) but lack an available recombination path. Below 50 K, there is no significant exciton thermalization that would result in a $\mathbf{\Gamma}$-centered recombination channel. At 100 K, $\mathbf{\Gamma}$-centered recombination begins to appear, predominantly driven by optical modes 7 and 9 (see Fig. 15-21 in the SI~\cite{Supplemental}). As the temperature increases, the lower-energy curves do not shift significantly but broaden, with satellites emerging in the energy range 20–40 meV above and below the bright line. The emergence of the $\mathbf{\Gamma}$-centered emission with increasing temperature, as shown in Fig.~\ref{fgr:fgr6}(a)-(c), indicates the redistribution of oscillator strength between indirect and direct excitons.  The dashed line, positioned $18~\mathrm{meV}$ below the bright exciton, marks the energy of the lowest-lying momentum-indirect (\textbf{K}-\textbf{K}$^\prime$) dark exciton. From this reference point, the threshold for LO phonon-assisted emission is located approximately $44.6~\mathrm{meV}$ away. Peaks appearing at energies lower (to the left) than this threshold arise predominantly from optical phonon-assisted transitions, as exemplified by the observed left peak corresponding to optical phonon mode 8. Conversely, peaks occurring at energies higher (to the right) are attributed primarily to acoustic phonon-assisted transitions, as illustrated by the right peak arising from acoustic phonon mode 3.\\
\noindent We now compare our results with that that in tungsten-based TMDs by Malic et al. \cite{Samuel2020}. It is reported that the bright exciton resonances dominate the PL at higher temperatures (e.g., 150 K and 300 K), leading to asymmetric emission spectra due to phonon-assisted recombination. In contrast, the recombination becomes more prominent at lower temperatures. Specifically, they found no observable phonon-assisted PL emission in unstrained Mo-based TMDs, consistent with our findings for unstrained MoS$_2$. However, under applied strain, our calculations in Fig.~\ref{fgr:fgr6} reveal prominent phonon-assisted PL features at lower temperatures. At higher temperatures (200 K and 300 K), the bright exciton resonance dominates again but exhibits asymmetry, consistent with observations in WSe$_2$ and WS$_2$. Notably, in strained MoS$_2$, phonon-assisted PL peaks appear approximately 50 and 70 meV below the bright exciton, aligning with reported shifts of 60 and 75 meV in WSe$_2$ and 70 and 90 meV in WS$_2$, respectively.
\noindent Malic et al. also noted a splitting of the phonon-assisted peak in hBN-encapsulated WSe$_2$ samples~\cite{Ye2018,Li2019,Courtade2017,Lindlau2017,Barbone2018}, revealing two distinct phonon-assisted emission features. A similar trend is observed in our calculations on strained MoS$_2$, predicting two peaks from different phonon modes: the peak closer to the bright exciton originates from the third phonon mode, while the more distant peak stems from the eighth mode.\\
\noindent To understand the shift in emission energy and the evolution of PL intensity with temperature, we present results under unstrained and +1$\%$ strain conditions in Fig.~\ref{fgr:fgr7}(a)-(b). A clear redshift of the bright emission is observed with increasing strain. Under unstrained conditions, faint indirect emission is visible near 1.73 eV, while the direct bright emission remains the strongest among all strain conditions. At lower strains (see Fig.~S25 in the SI~\cite{Supplemental}), the direct path intensities become most prominent due to satellite weights that thermalize at higher temperatures. The maximum intensity of the indirect excitons at cryogenic temperatures is observed at +1$\%$ strain. At higher strain levels, the indirect emissions become more distributed, but their intensities drop due to weaker exciton-phonon couplings. The appearance of doublet peaks at higher strains (beyond +1$\%$) reverses their intensity magnitude, as shown in Fig.~\ref{fgr:fgr5}, due to the additive contribution from other low-energy modes. Additional work remains to identify the symmetries of the indirect spin-split excitons $\mathrm{e}_{\mathbf{K}}$ with respect to C$_3$ rotation, which will be addressed in future studies.\\
\noindent We now discuss the limitation of our work. The formalism of $exc$-$ph$ coupling in Eq. (\ref{eqn:eqn2}) must be invariant if an arbitrary phase is added to the electronic wave functions entering the electron-phonon coupling (Eq. (\ref{eqn:eqn1})) and the exciton coefficients (A-terms) in Eq. (\ref{eqn:eqn2}). To preserve this gauge invariance, the same Kohn-Sham electronic wave function set $\left|n\mathbf{k}\right\rangle $ should be used, which are defined up to an arbitrary phase fixed by the specific simulation, for both the DFPT and BS steps. Otherwise, joining the \textit{g}-elements with the A-coefficient as complex numbers, like in Eq. (\ref{eqn:eqn2}), will lead to numerical noise and unphysical effects. This is a typical phase-mismatch which generates error both in the intensity peak magnitude as well as exciton lifetime. To the best of our knowledge, with the recent codes of Quantum Espresso and Yambo, this is impossible to do. For comparison, we highlight the works of Zanfrognini et al. \cite{PaleariVarsano} and Bernardi et al. \cite{Chen2020}, where the PERTURBO code \cite{zhou2021perturbo}, when combined with YAMBO, successfully reproduces the correct intensity levels corresponding to experimental peaks for bulk h-BN. Due to the complexity of interfacing these codes, our procedure does not utilize the YAMBO-PERTURBO interface. Thus, in this case, there could be a phase-mismatch which can show an opposite trend in the intensity levels, similar to Lechifflart et al. \cite{Lechifflart2023}. We have compared our exciton-phonon coupling results for the lowest-lying dark exciton with those of Chan et al. \cite{Chan2023} (see Fig. 2(b) in the SI\cite{Supplemental}). We find that, apart from the alternating patterns around the \textbf{M} point and the anomalous emergence of high coupling strength near the $\Gamma$ point in our Fig. \ref{fgr:fgr2} (which we attribute to phase-mismatch), our results are in strong agreement. A similar observation holds for the lowest-lying bright exciton (A exciton).\\
\noindent In summary, we investigate ab-initio exciton-phonon couplings in 2D MoS$_2$ using a quantum superposition approach for electron and hole scattering events with phonons. Through rigorous ab-initio calculations and group-theoretic analyses, we demonstrate that indirect emission in 2D MoS$_2$ under in-plane light polarization is symmetry-allowed for both acoustic and optical phonon modes with A$^{\prime}$ character. This systematic study serves as a textbook example of how indirect exciton recombination, absent in the compressive and unstrained cases, becomes dominated by hole-phonon scattering processes (\textbf{K} to \textbf{K}$^\prime$ transitions) under tensile strain. We find that this transition maximizes the indirect emission around +1$\%$ strain at cryogenic temperatures, driven by enhanced exciton-phonon scatterings assisted by intervalley A$^{\prime}$ phonon modes with reduced point group C$_S$(m). These results provide an alternative explanation for the non-monotonic PL variations with strain reported in recent experiments \cite{Pablo2022}.
% \begin{acknowledgments}
% \vspace{-1.2em}
 \\ \noindent This work was carried out with financial support from SERB, India with grant number CRG/2023/000476. We acknowledge the National Super Computing Mission (NSM) for providing computing resources of ``Paramshivay'' at IIT BHU, India. 
% \end{acknowledgments}
\bibliography{prb}

\end{document}